
\magnification=\magstep1
\hsize=128mm
\baselineskip=18pt
\vglue 1in

\centerline{\bf UNIVERSAL PARAMETRIC CORRELATIONS IN THE TRANSMISSION}
\centerline{\bf EIGENVALUE SPECTRA OF DISORDERED CONDUCTORS }
\vskip 10mm
\centerline{A. M. S. Mac\^edo }
\centerline{\it Theoretical Physics, University of Oxford,}
\centerline{\it 1 Keble Road, Oxford OX1 3NP, U.K.}
\vskip 10mm
\centerline{\bf Abstract}
\medskip
We study the response of the transmission eigenvalue spectrum of disordered
metallic conductors to an arbitrary external perturbation. For systems without
time-reversal symmetry we find an exact non-perturbative solution for the
two-point correlation function, which exhibits a new kind of universal behavior
characteristic of disordered conductors. Systems with orthogonal and symplectic
symmetries are studied in the  hydrodynamic regime.
\medskip

\noindent
PACS numbers: 72.10.Bg, 05.60.+w, 05.40.+j
\vfill\eject
Since its first application to the statistics of level spacings in the energy
spectra of complex nuclei [1], the random matrix theory of Wigner, Dyson, Mehta
and others has been developed into a powerful non-perturbative method to
describe a number of local statistical properties of many complex physical
systems, such as disordered metallic grains [2], quantum chaotic systems [3],
random surfaces [4] and disordered conductors [5].
Random matrix ensembles can be conveniently derived, for any given spectral
density, by means of a maximum-entropy requirement, which in turn implies that
the local spectral fluctuations are determined solely by some universal
symmetries of the system characterized by the parameter $\beta$: $\beta=1$ for
systems with time-reversal symmetry and without spin-orbit scattering,
$\beta=2$ for systems without time-reversal symmetry and $\beta=4$ for systems
with time-reversal symmetry in the presence of spin-orbit coupling. It is
precisely this universal behavior that accounts for the success of the many
applications of random matrix theory.

It is often difficult to find a microscopic justification for the
maximum-entropy ansatz that generates the random matrix ensembles. For the
particular case of disordered metallic grains, however, Efetov [6] has derived
a microscopic theory which, in the limit where a particle can diffuse
throughout the sample, yields precisely the same two-point correlation function
for pairs of energy levels obtained by random-matrix theory.

Very recently, Altshuler, Szafer and Simons [7] have extended Efetov's theory
to study spectrum correlations of disordered metallic grains in the presence of
an arbitrary external perturbation.
The most striking conclusion of their work is that the two-point correlation
function for pairs of energy levels at different values of the external
perturbation becomes {\it universal} if the energy levels and the perturbation
parameter are appropriately rescaled. In the hydrodynamic regime the
"velocity-velocity" correlator $C(\delta u)$ [7] has been shown to have a
remarkably simple form
$$
C(\delta u)=-{2 \over \pi^2 \beta \delta u^2},
\eqno (1)
$$
where $\delta u$ is the perturbation parameter. Going beyond the hydrodynamic
regime, Simons and Altshuler [7] have derived an exact non-perturbative
expression for $C(\delta u)$, which after appropriate rescaling of the physical
parameters becomes a universal function which is believed to be a new quantum
mechanical characterization of systems that obey the Wigner-Dyson statistics
[8] of consecutive level spacings.

Subsequently, Simons, Lee and Altshuler [9] have uncovered a number of
correlations between these results and other problems. In particular, they have
 shown that $C(\delta u)$ can also be derived from the periodic Sutherland
Hamiltonian [10], which is known to be equivalent to the Dyson Brownian motion
model [10].
The case $\beta=2$ is particularly simple, since the interaction term in the
Sutherland model vanishes and an exact solution for the two-point correlation
function has been calculated. A detailed account of the application of Dyson
Brownian motion model to describe parametric correlations in the spectra of
disordered metals has been given by Beenakker and Rejaei [11].

The universality discussed by the above authors occurs when the eigenvalues of
the random matrix are rescaled at the bulk of the spectrum.  However, large
random matrices are known to have eigenvalue densities limited to a finite
support and it has been found [12,13] that, in the absence of external
perturbation, different kinds of universal behavior emerge depending whether
the eigenvalues are rescaled at the endpoints of the support or at the bulk of
the spectrum.
It is therefore natural to expect a similar scenario when an adiabatic external
perturbation on the eigenvalues is turned on. The purpose of this letter is to
demonstrate that this is indeed what happens. In particular, we show that the
parameter dependent two-point correlation function for pairs of eigenvalues
belonging to the Laguerre ensemble gives rise to a new universal behavior when
rescaling takes place at the hard edge of the support.
Our main result (Eq. (4)) is a direct consequence of the Bessel kernel [13] in
the same way as the result of Refs. 7, 9 and 11, for $\beta=2$, is a direct
consequence  of the sine kernel. Since the Bessel kernel is a local property
which is known to appear at the hard edge of the support of the eigenvalue
density of many random matrix ensembles, we believe that Eq. (4) is of equal
general validity.

 The Laguerre ensemble has been applied to problems in several areas of
theoretical physics ranging from nuclear physics [14], QCD [13] and disordered
conductors [5]. In this letter we shall be mostly concerned with the random
matrix theory of disordered conductors.
The fundamental difference between the random matrix ensembles appropriate to
the description of disordered conductors and the more familiar gaussian
ensembles, used for modelling disordered metallic grains, is that one is not
concerned with the energy levels, $-\infty < E_i < \infty$, of a random
Hamiltonian, but rather with the parameters $\lambda_i \equiv 1/T_i-1$, where
$T_i$ are the eigenvalues of $tt^{\dagger}$ ($t$ is the transmission matrix),
which because of flux conservation must satisfy $0 \le \lambda_i < \infty$.
The positivity constraint on the $\lambda_i$'s implies that the two-point
correlation function is not translation invariant. Consequently, the variance
of a general linear statistic [15] differs from that for gaussian ensembles,
although the universal $1/\beta$ dependence is the same in both cases.

 It is interesting to compare random matrix theory with the microscopic theory
of disordered mesoscopic conductors  that has been developed by Lee and Stone
and by Altshuler [16]. In this theory one is interested in mesoscopic
fluctuations of the conductance, $G$, with the Fermi energy , $E_F$, or an
applied magnetic field , $B$.
This information is contained in the correlator $F(\delta u)=\langle G(\delta
u)G(0)\rangle-\langle G(0)\rangle^2$, where $\langle \dots \rangle$ denotes
ensemble average and $\delta u$ is a small perturbation on $E_F$ or $B$.
$F(\delta u)$ has been shown to decay as a power law as a function of $\delta u
/ {\cal E}_c$, where ${\cal E}_c$ is a correlation parameter which sets the
typical scale of the spacings between the peaks and valleys in the values of
$G$, for a given realization, as a function of $\delta u$.
This slow power law decay of $F(\delta u)$ indicates that the system sustains
some sort of long-range memory in the fluctuations of the conductance. This
kind of behavior must be contrasted with the exponential decay of Poisson
processes and can be regarded as a manifestation of "spectral rigidity", a well
known phenomenon in the theory of random matrices.

For the sake of mathematical tractability we shall consider the global
maximum-entropy approach [5]. This approach has been shown [17] to be an exact
description of the local statistical properties of levels that are close to
zero on scales comparable with the mean level spacing.
For levels away from the origin this approach provides an adiabatic
approximation for the actual distribution which turns out to be an excelent
approximation in the hydrodynamic limit. In the metallic regime [5] one has the
following probability distribution for the variables $x_i\approx
\sqrt{\lambda_i}$
$$
P^{\alpha}_0(\{x_i\})=C^\alpha_N \prod_{i<j} \left|x_i^2-x_j^2\right|^\beta
\prod_{i=1}^N x_i^{\beta \alpha+1} \exp(-{\beta \over 2} c x_i^2),
\eqno (2)
$$
where $c$ is a system dependent constant and $\alpha$ is a parameter that as we
shall demonstrate below labels different universality classes for the two-point
correlation function ($\alpha=0$ for disordered conductors). The most general
case, where $\alpha > -1$, was first derived by Bronk [14] and characterizes
the Laguerre ensemble of random matrices.

We now state our results. Assuming that spectrum correlations are dominated by
level repulsion, we propose a Brownian motion model (to be described below) for
the evolution, with an external perturbation parameter $\delta u$, of the
eigenvalues $\{x_i\}$, which is consistent with the asymptotic "equilibrium"
distribution $P^{\alpha}_0(\{x_i\})$. For $\beta=2$ we show that the resulting
Fokker-Planck equation can be mapped exactly onto a free-fermion Hamiltonian.
We calculate exactly the two-point correlation function defined as
$$
S(x,x',\delta u)\equiv \Bigl\langle \sum_{ij} \delta(x-x_i(u))
\delta(x'-x_j(u+\delta u)) \Bigr\rangle_{eq}-\rho(x)\rho(x'),
\eqno (3)
$$
where $\langle \dots \rangle_{eq}$ stands for an arbitrary average calculated
with the equilibrium distribution $P^{\alpha}_0(\{x_i\})$ and
$\rho(x)\equiv\langle \sum_i \delta(x-x_i(u))\rangle_{eq}$. After rescaling the
eigenvalues at the hard edge of the spectrum we find
$$\eqalignno{
S(x,x',\delta u)=&\pi^4\rho_0^4 \int_0^1 ds\int_1^\infty ds' ss'xx'
J_\alpha(\pi \rho_0 x s)J_\alpha(\pi \rho_0 x' s)J_\alpha(\pi \rho_0 x s') \cr
&J_\alpha(\pi \rho_0 x' s')\exp({\pi^2\rho_0^2 \over 2 \gamma}\delta u^2(s^2-
{s'}^2)),
& (4) \cr
}
$$
in which $J_\alpha(x)$ are the Bessel functions, $\rho_0$ is related to the
average density $\rho(x)$ through
$\rho(x)=\pi^2\rho_0^2\int_0^1 ds s x J_\alpha^2(\pi \rho_0 x s)$ and
$\gamma^{-1} = (\beta/2)\langle \dot x^2_i(u) \rangle$, where $\dot x_i \equiv
dx_i/du$, is the friction coefficient of the Brownian motion model. Note that
our expressions for $\rho(x)$ and $S(x,x',0)$ agree with known results [13].
The correlation parameter ${\cal E}_c$ discussed in the introduction must have,
by definition, the same dimension as $\delta u$. Since $x_i$ and $\rho_0$ are
dimensionless it is natural to introduce the definition ${\cal E}_c^{-2}=\beta
\rho_0 \langle \dot x^2(u) \rangle/2$. A straightforward evaluation yields
${\cal E}_c=\sqrt{\gamma/\rho_0}$.

Consider now the particular case of disordered conductors ($\alpha=0$).
Following Ref. 7  we make the rescalings $\hat x=\rho_0 x$, $\hat x'=\rho_0
x'$, $\delta \hat u^2=\delta u^2 \rho_0^2/\gamma$ and $\hat S(\hat x,\hat
x',\delta \hat u)=S(x,x',\delta u)/\rho_0^2$. We can see that $\hat S(\hat
x,\hat x',\delta \hat u)$ is a function that is independent of any physical
parameter, such as the Fermi velocity, the mean free path or the length of the
sample, and therefore can be regarded as a universal characterization of such
systems.
Since we started from the global maximum-entropy ansatz, Eq. (4) is exact for
disordered conductors only for levels that are close to the origin on scales of
the order of the mean level spacing. For the more general case of the Laguerre
ensemble, with $\alpha > -1$, we can see that the parameter $\alpha$ labels new
universality classes characteristic of systems with a hard edge in the
spectrum.

The "velocity-velocity" correlator defined as
$$
C(x,x',\delta u)\equiv \Bigl\langle \sum_{ij} \dot x_i(u) \dot x_j(u+\delta
u)\delta(x-x_i(u))
\delta(x'-x_j(u+\delta u)) \Bigr\rangle_{eq},
\eqno (5)
$$
is related to (4) by means of the identity $\partial_x\partial_{x'}C(x,x',u) =
-\partial_u^2 S(x,x', u)$. We shall consider $\langle \dot x_i \rangle_{eq}=0$
without loss of generality.

In the hydrodynamic regime fluctuations in the level density are small [18].
Using the fact that $\rho(x) \approx \rho_0$ for $x \gg 1/\rho_0$, we get
$C(x,x,\delta u)\approx -2/(\beta \pi^2 \delta u^2)$,
which is valid for arbitrary $\beta$ and $\delta u \gg u_c$. This is {\it
precisely} the universal correlator, $C(\delta u)$ of Eq. (1), found in Ref. 7.
 We can estimate $u_c$ if we recall that the integral of $C(x,x,\delta u)$ over
all $\delta u$ vanishes by definition, and use the definition of $\gamma$ to
get $\gamma^{-1}\rho_0^2 u_c \approx -\int_{u_c}^{\infty} C(x,x,\delta u)
d\delta u$.
Evaluating the integral we find $u_c^2 \approx \gamma / \rho_0^2$. For $\delta
u < u_c$ fluctuations in the level density are not small and  $C(x,x,\delta u)$
becomes, in general, an $\alpha$-dependent function.

We shall now only outline our calculations. The mathematical details will be
presented elsewhere [19]. Following Dyson's idea [10], we model the parametric
dependence of the distribution $P(\{x_i\},\tau)$ of the eigenvalues $\{ x_i \}$
by Brownian motion of $N$ classical particles at positions $\{x_i(\tau)\}$
moving in a viscous fluid with friction coefficient $\gamma^{-1}$ at
temperature $\beta^{-1}$. The parameter $\tau$ is a fictitious "time", which is
related to the perturbation parameter $\delta u$ by $\tau=\delta u^2$ [9,11].
We require the initial condition $P(\{x_i\},0)=\prod_i \delta(x_i-x_i^0)$,
where $x_i^0$ are the transmission eigenvalues of a perturbation-free system.
We also require that as $\tau \to \infty$ the system tends to a maximum-entropy
configuration, therefore $P(\{x_i\},\infty)=P^{\alpha}_0(\{x_i\})$. The
simplest description of a Brownian motion satisfying these requirements is
given by the Fokker-Planck equation
$$
\gamma{\partial P\over \partial \tau} =
\sum_{i=1}^{N} {\partial \over \partial x_i}
\left(P{\partial W \over \partial x_i}+{1\over \beta}{\partial P \over \partial
x_i}\right),
\eqno (6a)
$$
where
$$
W(\{x_i\})=-(\alpha+{1 \over \beta})\sum_i \ln x_i + {c \over 2} \sum_i x_i^2
-\sum_{i<j} \ln\left|x_i^2-x_j^2\right|.
\eqno (6b)
$$
More general descriptions can be obtained by allowing for other sources of
correlations in addition to the pure geometric effect of level repulsion (see
Ref. 11 for a short discussion).
As in the case of the conventional Dyson Brownian motion model we shall ignore
these additional effects and assume that correlations are dominated by level
repulsion.

We proceed by writing (6) and (2) in terms of the variables $\lambda_i = x_i^2$
and then we perform the transformation $\hat P(\{\lambda_i\},\tau)=\hat
P^{\alpha}_0(\{\lambda_i\})^{1/2} \Psi(\{\lambda_i\},\tau)$.
We find that $\Psi(\{\lambda_i\},\tau)$ satisfies the Schr\"odinger equation in
imaginary time $-\partial \Psi/\partial \tau=({\cal H}-B)\Psi$,
where
$$
{\cal H}=\omega
 \sum_i \biggl[
-{\partial \over \partial r_i}r_i{\partial \over \partial r_i}+\left({\beta
\over 4}\right)^2(r_i+{\alpha^2 \over r_i})+
{\beta(\beta-2) \over 4}\sum_{j(\ne i)} {r_i \over (r_i-r_j)^2} \biggr ],
\eqno (7)
$$
$B=N c(2+\beta(N+\alpha-1))/(2\gamma)$, $\omega=4c/(\beta \gamma)$ and
$r_i=c\lambda_i$. Observe that for $\beta=2$, ${\cal H}$ becomes a
non-interacting Hamiltonian, whose single electron eigenfunctions are the
Laguerre functions $\Phi_n^\alpha(\lambda)$. Using its asymptotics and taking
$N \to \infty$ and $c \to 0$, such that $2\sqrt{Nc} \to \pi \rho_0$, we can
evaluate $S(x,x',\delta u)$ exactly to find Eq. (4). A more detailed discussion
of $S(x,x',\delta u)$ can be found in Ref. 19. We now discuss the hydrodynamic
regime.

The hydrodynamic limit can be calculated directly from (6) for arbitrary
$\beta$. Multiplying (6) by $\sum_i\delta(x-x_i)$ and integrating over all
$x_i$'s we get the following evolution equation for the average density
$$
\gamma {\partial \rho(x,\tau) \over \partial \tau} \approx
-{\partial \over \partial x}\left[ \rho(x,\tau) {\partial \over \partial x}
\left( \int_0^\infty dy\rho(y,\tau)\ln\left|x^2- y^2\right|-{c \over
2}x^2\right)\right],
\eqno (8)
$$
in which only the dominant contributions for large $N$ were considered. The
equilibrium density is $\rho_{eq}(x)\approx 2\pi^{-1}\sqrt{Nc} \to \rho_0=
const$, and thus tends to a constant once the limiting procedure introduced
above is implemented.

We are concerned with the diffusive regime, where the average dimensionless
conductance, $\langle G \rangle_{eq}$, satisfies $\langle G \rangle_{eq} \gg
{\rm var}(G) \sim O(1)$. Since according to the Landauer formula the average
conductance is simply $\langle G \rangle_{eq} \approx \rho_0$, one expects
fluctuations in $\rho_{eq}(x)$ to be small in this regime.
Defining $\delta \rho(x,\tau)\equiv \rho(x,\tau)-\rho_0$, inserting into (8),
neglecting contributions of second order in $\delta \rho$ and taking the
Fourier cosine transform of the resulting equation, we find $\delta \tilde
\rho(k,\delta u^2) =\delta \tilde \rho(k,0)\exp(-k \pi \delta u^2 /{\cal
E}_c^2)$.
 From (3) we get for the Fourier transform of $S(x,x',\delta u)$ the result
$\tilde S(k,\delta u)=k/(2\beta) \exp(-k \pi \delta u^2/ {\cal E}_c^2)$,
in which we have used Beenakker's result [15] for $\tilde S(k,0)$. This
equation is valid in the range $k \ll \rho_0$ for all $\delta u$.

We are now in a position to calculate $C(x,x,\delta u)$. Using the relation
between $C(x,x',u)$ and  $S(x,x', u)$ and the above expression for $\tilde
S(k,\delta u)$ we find $C(x,x,\delta u)\approx -2/(\beta \pi^2 \delta u^2)$,
for $\delta u \gg u_c$ and $x \gg 1/\rho_0$, this result is just Eq. (1).

Finally, we shall use our result for $S(x,x',\delta u)$ to calculate parametric
correlations of the conductance. Defining  $\delta G \equiv G-\langle G \rangle
_{eq}$ and $F(\delta u)\equiv
\langle \delta G(\delta u) \delta G(0) \rangle_{eq}$ we find
$$
F(\delta u)= {1 \over 8 \beta}\left(1+{\pi\over 2}{ \delta u^2 \over {\cal
E}_c^2}\right)^{-2}.
\eqno (9)
$$
Observe that when $\delta u =0$, Eq (9) gives var$(G)=1/(8\beta)$, which is the
value found in Ref. 15.  For $\delta u \ll {\cal E}_c$ one gets from (9)
$F(\delta u) \approx 1/(8\beta)(1-\pi\delta u^2/{\cal E}^2_c)$, a quadratic
decrease, as found in Ref. 16. For $\delta u \gg {\cal E}_c$ we find $F(\delta
u) \approx (\delta u/{\cal E}_c)^{-4}$, which is a power law decay as one would
have expected from spectral rigidity and is also in qualitative agreement with
Ref. 16.
The exact value of the exponent found in Ref. 16 shows a nontrivial dependence
on both the dimensionality and the nature of the perturbation, which cannot be
obtained from the present random matrix model. The overall shape of $F(\delta
u)$ in Eq. (9), therefore, is in reasonable good qualitative agreement with
microscopic calculations.

In conclusion we have presented a consistent theory of parametric correlations
in the Laguerre ensemble. As a particular case we have considered the problem
of the response to an arbitrary external perturbation of the transmission
eigenvalue spectrum of disordered metallic conductors. We believe that some of
our intermediate results, like Eqs. (6) and (7), might be of general interest
in connection with applications of the Laguerre ensemble in other branches of
physics.

The author would like to thank J. T. Chalker for many useful comments and
illuminating discussions. I am especially grateful to B. D. Simons for a
careful reading of the manuscript and many useful suggestions. This work was
partially supported by CAPES (Brazilian Agency).

\bigskip
\noindent
{\bf References}
\medskip
{\parindent=0.5cm
\item{[1]} See e.g. C. E. Porter, {\it Statistical Theory of Spectra:
Fluctuations} (Academic, New York, 1965).
\item{[2]} L. P. Gorkov and G. M. Eliashberg, Zh. Eksp. Teor. Fiz. {\bf 48},
1407 (1965) [Sov. Phys. JETP {\bf 21}, 940 (1965)].
\item{[3]} See e.g. F. Haake, {\it Quantum Signatures of Chaos} (Springer,
Berlin, 1992).
\item{[4]} E. Br\'ezin, C. Itzykson, G. Parisi and J. B. Zuber, Commun. Math.
Phys. {\bf 59}, 35 (1978).
\item{[5]} For a review see A. D. Stone, P. A. Mello, K. A. Muttalib, and J.-L.
Pichard, in {\it Mesoscopic Phenomena in Solids}, edited by B. L. Altshuler, P.
A. Lee, and R. A. Webb (North-Holland, Amsterdam, 1991)
\item{[6]} K. B. Efetov, Adv. Phys. {\bf 32}, 53 (1983).
\item{[7]} A. Szafer and B. L. Altshuler, Phys. Rev. Lett. {\bf 70}, 587
(1993); B. D. Simons and B. L. Altshuler, ibid. {\bf 70}, 4063 (1993); Phys.
Rev. B {\bf 48}, 5422 (1993).
\item{[8]} E. P. Wigner, Ann. Math. {\bf 53}, 36 (1951); F. J. Dyson, J. Math.
Phys. {\bf 3}, 140 (1962).
\item{[9]} B. D. Simons, P. A. Lee and B. L. Altshuler, Nucl. Phys. {\bf B409},
487 (1993); Phys. Rev. Lett. {\bf 70}, 4122 (1993).
\item{[10]} B. Sutherland, J. Math. Phys. {\bf 12}, 246 (1971); {\bf 12}, 251
(1971); Phys. Rev. A {\bf 4}, 2019 (1971); {\bf 5}, 1372 (1972); F. J. Dyson,
J. Math. Phys. {\bf 13}, 90 (1972).
\item{[11]} C. W. J. Beenakker, Phys. Rev. Lett. {\bf 70}, 4126 (1993); C. W.
J. Beenakker and B. Rejaei Physica A {\bf 203}, 61 (1994).
\item{[12]} M. J. Bowick and E. Br\'ezin, Phys. Lett. B {\bf 268}, 21 (1991);
D. Fox and P. B. Kahn, Phys. Rev. B {\bf 134}, 1151 (1964).
\item{[13]} J. J. M. Verbaarschot and I. Zahed, Phys. Rev. Lett. {\bf 70}, 3852
(1993); P. J. Forrester, Nucl. Phys. {\bf B402}, 709 (1993).
\item{[14]} B. V. Bronk, J. Math. Phys. {\bf 6}, 228 (1965).
\item{[15]} C. W. J. Beenakker, Phys. Rev. Lett. {\bf 70}, 1155 (1993).
\item{[16]} P. A. Lee and A. D. Stone, Phys. Rev. Lett. {\bf 55}, 1622 (1985);
B. L. Altshuler, Pis'ma Zh. Eksp. Teor. Fiz. {\bf 41}, 530 (1985)[JEPT Lett.
{\bf 41}, 648 (1985)]; P. A. Lee, A. D. Stone, and H. Fukuyama, Phys. Rev. B
{\bf 35}, 1039 (1987).
\item{[17]} C. W. J. Beenakker and B. Rejaei, Phys. Rev. Lett. {\bf 71}, 3689
(1993) ; J. T. Chalker and A. M. S. Mac\^ edo, ibid. {\bf 71}, 3693 (1993).
\item{[18]} It is important to bear in mind that although the hydrodynamic
limit constitutes only an adiabatic approximation [17] to the problem of
transmission eigenvalues ($\alpha=0$), we find that since the final results do
not depend on $\alpha$ it must be of considerable general interest to the
reader concerned with other applications of random matrix theory, where the
Laguerre ensemble description is exact [13,14].
\item{[19]} A. M. S. Mac\^ edo (unpublished).

\bye